\documentclass[10pt]{iopart}
\usepackage{epsfig,graphicx}

\begin{document}

\title{Inhomogeneous Kibble-Zurek mechanism: vortex nucleation during Bose-Einstein condensation}

\author{A. del Campo$^{1,2}$, A. Retzker$^{2}$, M. B. Plenio$^{2}$}
\address{$^1$ Institut f{\"u}r Theoretische Physik, Leibniz Universit\"at Hannover, Appelstr. 2 D-30167,
Hannover, Germany}

\address{$^2$ Institut f{\"u}r Theoretische Physik, Albert-Einstein Allee 11,
Universit{\"a}t Ulm, D-89069 Ulm, Germany}

\newcommand{\beqa}{\begin{eqnarray}}
\newcommand{\eeqa}{\end{eqnarray}}

\newcommand{\om}{\omega}

\begin{abstract}
The  Kibble-Zurek mechanism is applied to the spontaneous  formation of vortices in a harmonically trapped thermal gas following a temperature quench through the critical value for Bose-Einstein condensation.  While in the homogeneous scenario vortex nucleation is always expected, we show that it can be completely suppressed in the presence of the confinement potential, whenever the speed of the spatial front undergoing condensation is lower than a threshold velocity. Otherwise, the interplay between the geometry and causality leads to different scaling laws for the density of vortices as a function of the quench rate, as we also illustrate for the case of a toroidal trapping potential.

\end{abstract}


\maketitle

\section{Introduction}
Non-equilibrium phase transitions generally lead to  phases with limited long-range order.
When a system is quenched through a critical point of a second-order phase transition both the correlation length $\xi$
and relaxation time $\tau$ diverge. At the freeze-out time,  $\hat{\mathrm{t}}$, the relaxation time of the system
equals the time scale of the quench, the dynamics essentially freezes (impulse stage), and there is a breakdown of adiabaticity. The paradigmatic Kibble-Zurek theory predicts that the average size of the domains in the low-symmetry phase
is given by the correlation length at the freeze-out time \cite{Kibble,Zurek}
\beqa
\hat{\xi}:=\xi(\hat{\mathrm{t}}).
\eeqa
Ultimately, the freeze-out time depends on the quench rate $1/\tau_Q$, which leads to a scaling law of the density of defects as a function of the rate at which the transition is crossed,
\beqa
\mathcal{D}=\hat{\xi}^{-D}\sim \tau_Q^{\alpha},
\eeqa
where $D$ is the dimension of the domains being considered and $\alpha<0$. It is well known that dissipation as well as other non-universal mechanisms for defect losses (such as annihilation by scattering of defects with opposite
topological charges) might lead to deviations from the Kibble-Zurek scaling whenever they become dominant. Nonetheless, this scenario is supported by different numerical studies \cite{kzmnum}, and experiments aimed at the confirmation of this prediction have been carried out in a variety of systems \cite{kzmexp}, see \cite{review} for a recent review.

Recent experiments with pancake shaped Bose-Einstein condensates have reported the spontaneous nucleation of vortices during condensation \cite{BrianAnderson08}.
Starting with a thermal gas in an oblate harmonic trap a linear quench in the temperature was applied to induce condensation.
Such scenario can be naturally discussed in the light of the
Kibble-Zurek mechanism \cite{Kibble,Zurek,Zurek96}.
A crucial feature in the experiments is the inhomogeneous character of the system arising from the external trapping potential.
As a consequence, the transition does not occur simultaneously in the entire system
and the homogeneous Kibble-Zurek mechanism (HKZM) described above breaks down, making it necessary to extend it to scenarios
where the nucleation of defects is governed by causality \cite{ikzm1,ikzm2,ikzm3,Zurek09bec,ikzm4,ikzm5,ikzm6}.
To date, there is no experimental evidence supporting this extension that we shall refer to as the Inhomogeneous Kibble-Zurek mechanism (IKZM),
and whose main prediction is the existence of two regimes: a) one in which the phase transition is crossed adiabatically with complete suppression of nucleation of defects
b) other, characterized by vortex nucleation, where the density of defects after the quench obeys a scaling law with the quenching rate,
different from that in the homogeneous mechanism described above and
governed by the inhomogeneities in the system.
Due to the high-control of the trapping potential which induces the inhomogeneous density profile in a trapped cloud,
the nucleation of vortices during Bose-Einstein condensation stands out as an ideal scenario to test the predictions of the IKZM
and it is highly desirable to extend the results of the experiments in \cite{BrianAnderson08} to such aim.
Here, we analyze theoretically in this experimental setup the benchmarks of the IKZM, which are key for its verification.

\section{The Homogeneous Kibble-Zurek mechanism}

We start recalling the results of the HKZM for a uniform thermal gas \cite{kzmvortex}.
Consider a uniform quench of the temperature $T(t)$ across the critical value of condensation $T_c$. For a symmetric linear quench between the initial $T_i=T_c(0)+\delta$ and final $T_f=T_c(0)-\delta$ temperatures, it follows that
\beqa
T(t) &=& T_i-t\frac{T_i-T_f}{\tau},\nonumber\\
&=&T_c(0)\left(1-\frac{\mathrm{t}}{\tau_Q}\right),
\eeqa
where $\mathrm{t}=t-\tau/2$ and
\beqa
\tau_Q=\tau\frac{ T_c(0)}{2\delta}.
\eeqa
The reduced temperature
\beqa
\epsilon(\mathrm{t})=\frac{T_c(0)-T(\mathrm{t})}{T_c(0)}
\eeqa
governs the divergence of both the correlation length
\beqa
\xi(\mathrm{t})=\frac{\xi_0}{|\epsilon(\mathrm{t})|^{\nu}}
\eeqa
and the relaxation time
\beqa
\tau(\mathrm{t})=\frac{\tau_0}{|\epsilon(\mathrm{t})|^{\nu z}}
\eeqa
as the system approaches the critical point ($\epsilon(\mathrm{t})=0$). Here, $\{z, \nu\}$ are the critical exponents  determined by the universality class to which the system belongs to. The instant $\hat{\mathrm{t}}$ in which the relaxation time equals the time remaining to the transition,
\beqa
\tau(\hat{\mathrm{t}})=\frac{\epsilon}{\dot{\epsilon}}\Big|_{\hat{\mathrm{t}}}=:\hat{\mathrm{t}},
\eeqa
is the {\it freeze-out time} which fixes the area $\hat{\xi}^2=\xi(\hat{\mathrm{t}})^2$ of the spots where the phase of the condensate is picked homogeneously.
It has been theoretically shown \cite{mergingtheory} and experimentally demonstrated \cite{Anderson07} that merging independent condensates with uniform random phases can lead to the nucleation of vortices. The geometrical configuration in the merging process  determines the yield according to the geodesic principle \cite{Berry}. The efficiency of this process can be captured by a constant $f$ independent of the critical exponents of the system. Moreover, since the system is homogeneous, so it is the transition, and defects might nucleate everywhere in the system. It follows that the density of vortices that spontaneously nucleate under such a quench can be estimated as the inverse of the square of the correlation length at the freeze out time,
\beqa
\label{HKZM}
\mathcal{D}_{\rm HKZM}=\frac{1}{f\hat{\xi}^2}=\frac{1}{f\xi_0^2}\left(\frac{\tau_02\delta}{\tau T_c(0)}\right)^{\frac{2\nu}{1+\nu z}}.
\eeqa
Experiments on the Bose-Einstein condensation of a 3D thermal cloud \cite{Donner07} have reported a critical exponents in agreement with the static 3D XY universality class, for which the best theoretical estimate to date is  $\nu = 0.6717(1)$ \cite{3DXY}. For our purposes the approximation $\nu\simeq 2/3$ will suffice.
The dynamic critical exponents is expected to be $z = 3/2$ as in the superfluid transition in $^4$He, the model F in the classification of Hohenberg and Halperin \cite{HH77}, up to possible small deviations discussed in \cite{FM06}. This leads to a dependence
\beqa
\mathcal{D}_{\rm HKZM}\sim\tau_Q^{-2/3},\quad {\rm while } \quad \mathcal{D}_{\rm HKZM}\sim\tau_Q^{-1/2}
\eeqa
follows from the mean-field values  $\nu = 1/2$, $z =2$.
Finite-size effects might pave the way to an adiabatic transition whenever the correlation length at the freeze-out time surpasses the size of the system.
Other than that, nucleation of vortices will take place no matter how slowly the temperature is quenched.

\section{The Inhomogeneous Kibble-Zurek mechanism}

In the following we focus on the role of the inhomogeneities arising as a result of the external trapping potential.
We shall see that its presence brings two new ingredients,
a local critical temperature and a local quench rate, changing the power-law for
the density of defects as a function of the quench rate.

Let us consider a thermal gas confined in an 3D oblate harmonic trap, isotropic in the radial direction, along which the density distribution exhibits a Gaussian profile of the form $n(r)=n_0e^{-\frac{m\om^2}{2k_BT}r^2}$.
And let us focus on the nucleation of vortices on the equatorial plane.
Due to the inhomogeneous density, the critical temperature acquires a dependence on the radial position \cite{Zurek09bec},
\beqa
T_c(r)=T_c(0)e^{-\frac{m\om^2}{3k_BT}r^2}.
\eeqa
Here, $T_c(0)$
is the critical temperature for the homogeneous system
 \cite{BK91,KvD96}.
For compactness, we shall introduce the thermal length $\Delta=\sqrt{3k_BT/2m\om^2}$.
Due to the spatial dependence of $T_c(r)$ , as the temperature is quenched different parts of the system undergo condensation at different times.
It turns out to be convenient to introduce the spatially-dependent reduced temperature
\beqa
\epsilon(r,\mathrm{t})=\frac{T_c(r)-T(\mathrm{t})}{T_c(r)},
\eeqa
to identify a front in the system crossing the transition at a given position and time
$(r_F,\mathrm{t}_F)$ satisfying the condition
\beqa
\epsilon(r_F,\mathrm{t}_F)=0.
\eeqa
It follows that $\frac{\mathrm{t}_F}{\tau_Q}=1-\frac{T_c(r)}{T_c(0)}$,
which in turn, allows us to rewrite the relative temperature as $\epsilon(r,\mathrm{t})=\frac{\mathrm{t}- \mathrm{t}_F}{\tau_Q(r)}$,
with a radial-dependent quench time
\beqa
\tau_Q(r)=\tau_Q\frac{T_c(r)}{T_c(0)}.
\eeqa
As a result, the relaxation time $\hat{\tau}=\hat{\tau}(r)$ and correlation length $\hat{\xi}=\hat{\xi}(r)$ at the freeze-out time acquire a local dependence.
The process resembles closely the formation of solitons in a 1D BEC \cite{Zurek09bec}.
The nucleation of topological defects in an inhomogeneous transition is governed by causality \cite{ikzm1,ikzm2,ikzm3,Zurek09bec,ikzm4,ikzm5,ikzm6}.
Indeed, when the front of the transition moves faster than the characteristic velocity of a perturbation, defects nucleate,
while otherwise the choice of the order parameter in the broken symmetry phase is done homogeneously along the system.

Approaching the transition from the high symmetry phase, the thermal gas starts to condense in the center 
of the cloud where the density and critical temperature are higher. The condensate, where the $U(1)$ symmetry is broken, grows then radially, the velocity of the front being
\beqa
v_F&=&\frac{T_c(0)}{\tau_Q}\left\vert\frac{dT_c(r)}{dr}\right\vert^{-1},\nonumber\\
&=&\frac{\Delta^2}{\vert r \vert \tau_Q(r)},
\eeqa
where we have disregarded corrections due to the time derivative of the critical temperature
$T_c(r)=T_c(r,\mathrm{t})$ arising from the quench $T(\mathrm{t})$ which affects the density profile.
This is reasonable whenever the amplitude of the quench is small $\delta\ll T_c(0)$.
The existence of this front and a local temperature play a crucial role in the following, and requires a local thermalization  fast compared to the time scale of the quench.
For defects to nucleate, the front of the transition is to move faster than any perturbation.
The characteristic velocity of a perturbation when the dynamics stops being adiabatic, this is, at the freeze-out time,
can be upper-bounded by the ratio of the freeze-out correlation length over the relaxation time,
\beqa
\hat{v}=\frac{\hat{\xi}}{\hat{\tau}}=\frac{\xi_0}{\tau_0}
\left(\frac{\tau_0}{\tau_Q(r)}\right)^{\frac{\nu(z-1)}{1+\nu z}}.
\eeqa
While within the HKZM the appearance of vortices is always expected (up to finite-size effects of the system), the trapping potential paves the way for
a perfectly adiabatic condensation, where the spontaneous nucleation is suppressed as long as
\beqa
v_F<\hat{v}.
\eeqa
Conversely, the condition for appearance of defects, $v_F>\hat{v}$, leads to a threshold value of the radius of the cloud delimiting the area
within which defects might nucleate,
\beqa
\label{r1}
\vert\hat{r}\vert&<&\frac{\Delta^2}{\xi_0}\left(\frac{\tau_0}{\tau_Q(r)}\right)^{\frac{1+\nu}{1+\nu z}},\nonumber\\
&<&\frac{\Delta^2}{\xi_0}\left(\frac{\tau_0}{\tau_Q}\right)^{\frac{1+\nu}{1+\nu z}}
\exp\left(\frac{\hat{r}^2}{2\Delta^2}\frac{1+\nu}{1+\nu z}\right).
\eeqa
Within a reduced area of the condensate  $\mathcal{S}_*$, formation of vortices occurs as in the HKZM, setting the average size  $\hat{\xi}^2$  of the spots with uniform local phase. Up to a numerical factor $f$,
the total number of topological defects cab be estimated as follows
\beqa
\mathcal{N}_{\rm IKZM}\simeq\int_{\{r|v_F>\hat{v}\}}dr\frac{2\pi r}{f\hat{\xi}(r)^2}
\eeqa
and simply by $\mathcal{N}_{\rm IKZM}=\frac{\mathcal{S}_*}{f\hat{\xi}^2}$ if one neglects the radial dependence of $\hat{\xi}$.

The transcendental inequality (\ref{r1}) can be approximated around the center of the cloud ($\hat{r}\ll\Delta$) by setting $\tau_Q(r)=\tau_Q(0)=\tau_Q$, and consistently $\hat{\xi}(r)=\hat{\xi}(0)$.
Solving the associate equality one finds the effective radius $\hat{r}=r_*$, in terms of which $\mathcal{S}_*=\pi r_*^2$
so that
\beqa
\label{IKZMscaling}
\mathcal{N}_{\rm IKZM}^{(0)}=\frac{\pi \Delta^4}{f \xi_0^4}
\left(\frac{\tau_02\delta}{\tau T_c(0)}\right)^{\frac{2(1+2\nu)}{1+\nu z}}.
\eeqa
We note that for the mean-field critical exponents, $z=2$ and $\nu=1/2$,
the exponent governing the scaling in Eq. (\ref{IKZMscaling}) is given by $\frac{2(1+2\nu)}{1+\nu z}=2$, which is
four times that predicted by the HKZM. For $z=3/2$ and $\nu=2/3$, the IKZM power-law exponent becomes $7/3$, which is as well nearly four times that of the homogeneous counterpart,
and constitutes an experimentally accessible benchmark of the IKZM.

Nonetheless, note that the absolute number of defects is reduced by a factor
\beqa
\mathcal{N}_{\rm IKZM}/\mathcal{N}_{\rm HKZM}\sim (r_*/r_M)^2\sim
\left(\frac{\Delta^2}{\xi_0r_M}\right)^2\left(\frac{\tau_0}{\tau_Q}\right)^{\frac{2(1+\nu)}{1+\nu z}},
\eeqa
 where $r_M$ is the radius of the cloud.
Hence, it is the dependence on the quench rate and the critical exponents  inherited by the effective size of the cloud $\mathcal{S}_*$ and $\hat{\xi}$ that is responsible for the new power-law governing the density of defects.

\begin{figure}
\begin{center}
\includegraphics[width=0.55\linewidth]{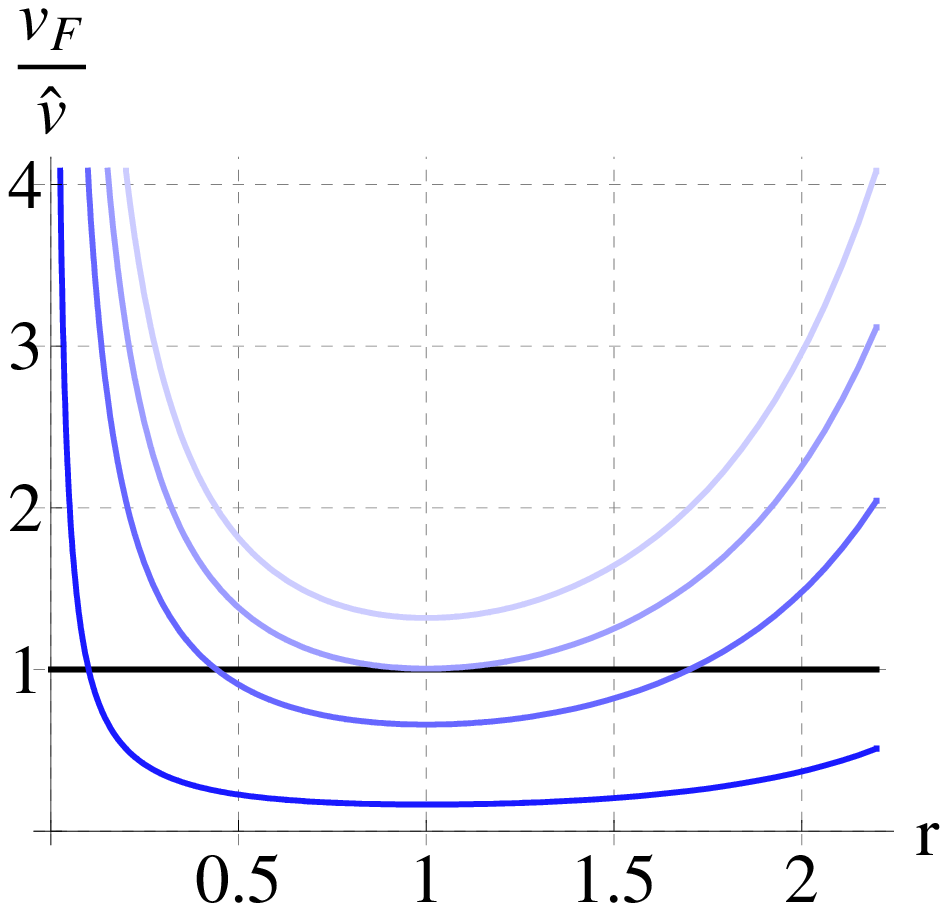}
\end{center}
\caption{\label{figvfvs}
Ratio of the velocity of the front $v_F$ crossing the critical point for condensation, and the characteristic speed of a perturbation $\hat{v}$ at the freeze-out time,
along the radial axis of a pancake-shaped atomic thermal cloud, for the values $\mathcal{A}=0.1, 0.4, 1/\sqrt{e}, 0.8$ (from bottom to top). The coefficient $\mathcal{A}=\sqrt{\frac{1+2\nu}{1+\nu z}}\Theta\Upsilon^{\frac{1+\nu}{1+\nu z}}$ allows one to
identify the homogeneous ($\mathcal{A}>1/\sqrt{e}$) and inhomogeneous ($\mathcal{A}\leq1/\sqrt{e}$) scenarios.
}
\end{figure}

For the general solution of Eq. (\ref{r1}), it is convenient to introduce the variable
\beqa
\zeta(r)=\frac{\hat{r}}{\Delta}\sqrt{\frac{1+2\nu}{1+\nu z}},
\eeqa
to find $\zeta<\mathcal{A} e^{\frac{\zeta^2}{2}}$,
with
\beqa
\mathcal{A}=\frac{\Delta}{\xi_0}\sqrt{\frac{1+2\nu}{1+\nu z}}\left(\frac{\tau_0}{\tau_Q}\right)^{\frac{1+\nu}{1+\nu z}}.
\eeqa
Figure \ref{figvfvs} shows the ratio $v_F/\hat{v}$ along the radial coordinate of the cloud for different values of this parameter.
In the central region where the density reaches its maximum, and hence it is approximately uniform,  this ratio is always larger than unity.

\begin{figure}
\begin{center}
\includegraphics[width=0.8\linewidth]{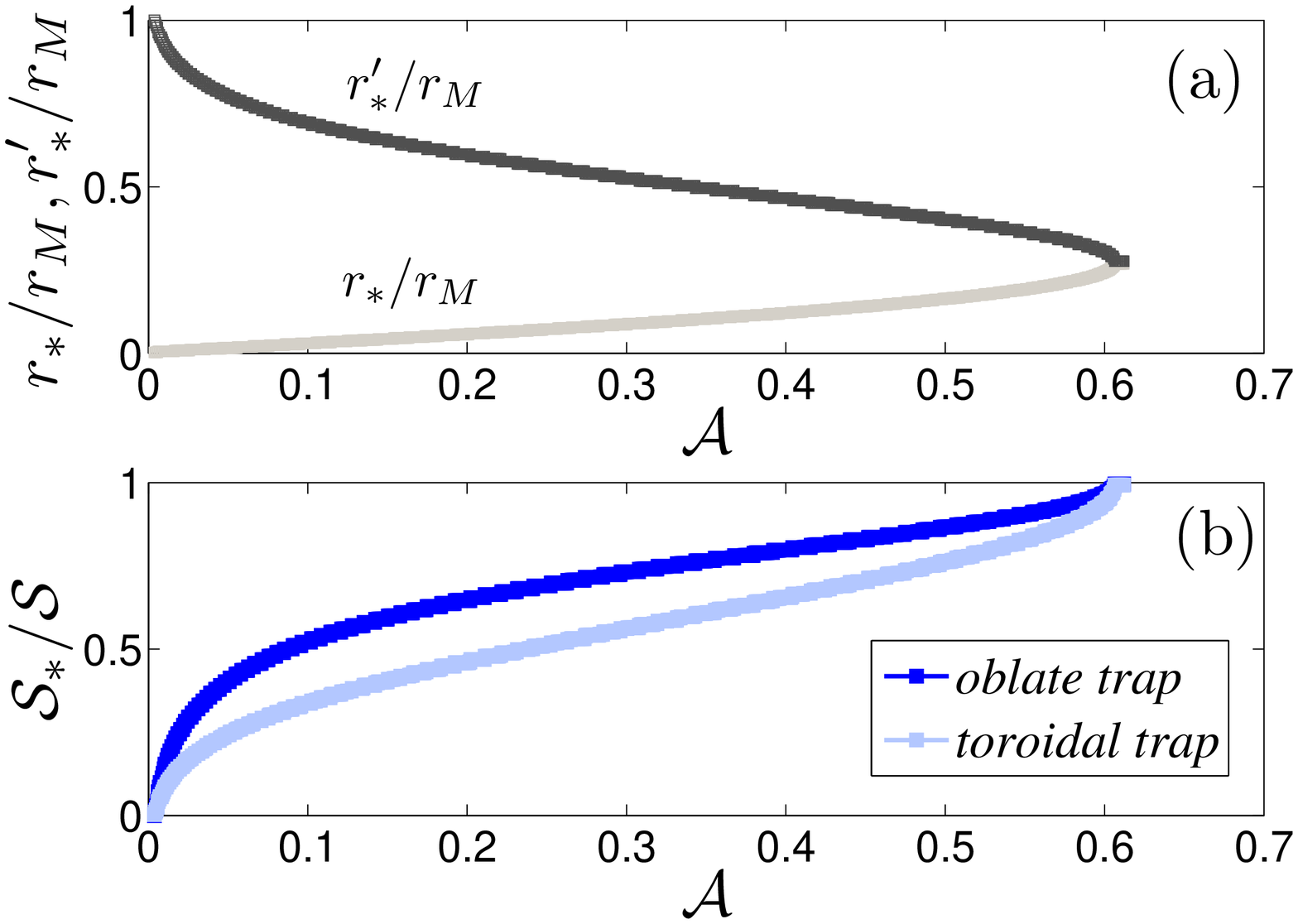}
\end{center}
\caption{\label{effsize}

Reduction due to causality of the effective size of the cloud for nucleation of vortices.
a) Critical values $r_*$ and $r_*'$ of the radial coordinate of the cloud determining the regions where defects might nucleate as a function of $\mathcal{A}<\mathcal{A}_c$.
b) Effective area as a function of the parameter $\mathcal{A}$ within the IKZM
for a pancake atomic cloud as well as for a cloud in a toroidal trap, which will be discussed in Section 4.
}
\end{figure}

%
One can distinguish two different regimes as a function of the value of $\mathcal{A}$
with respect to the critical value $\mathcal{A}_c=1/\sqrt{e}$:
 (i) For $\mathcal{A}\geq\mathcal{A}_c$, $v_F$ is everywhere along the sample larger than $\hat{v}$, so the homogeneous KZM applies. The density of vortices is then  given by Eq. (\ref{HKZM}).
(ii) For $\mathcal{A}<\mathcal{A}_c$ there exist two solutions $\{\zeta_*=\zeta(r_*)<\zeta_*'=\zeta(r_*')\}$  defining two disjoint concentric disks with support on the interval  $\Gamma:=[0,r_*]\cup[r_*',r_M]$ along the radial direction, where  $r_M$ is the effective radius of the cloud. As a result, see Fig. \ref{effsize}, the area $\mathcal{S}_*(\mathcal{A})=\pi[r_*^2+(r_M^2-r_*'^2)]$ where vortices might nucleate is reduced with respect the total area of the cloud $\mathcal{S}=\pi r_M^2$, and so it is the corresponding density of defects (note that case (i) corresponds to $r_*=r_*'$, so the effective area $\mathcal{S}_*$ equals the total area $\mathcal{S}$).
 Nonetheless, for $\mathcal{A}<\mathcal{A}_M=\zeta(r_M) e^{-\frac{\zeta(r_M)^2}{2}}$,  $r_*'>r_M$ so the outer disk can be ignored, and nucleation can be considered to take place only in the center of the cloud ($r<r_*$). This is the regime where Eq. (\ref{IKZMscaling}) holds, up to finite size effects which might lead to an adiabatic transition and a breakdown of the scaling whenever $\hat{\xi}>r_*$.
These different cases follow from the spatial distribution of the regions where the  ratio $v_F/\hat{v}>1$ as a function of $\mathcal{A}$, exhibiting a transition from case (i) to (ii). For instance, if in a given experiment only the quenching time is varied, one can identified a critical value
\beqa
\tau_Q({\mathcal A}_c)=\tau_0\Bigg[\frac{\xi_0}{\sqrt{e}\Delta}\sqrt{\frac{1+\nu z}{1+2\nu}}\Bigg]^{\frac{1+\nu z}{1+\nu}}
\eeqa
around which the scaling changes between those predicted by HKZM and IKZM.

To appreciate how the scaling is modified, we introduce the  dimensionless cloud width and quench rate
\beqa
\Theta=\Delta/\xi_0, \qquad  \Upsilon=\tau_0/\tau_Q,
\eeqa
and rewrite the parameter $\mathcal{A}$ as
\beqa
\mathcal{A}=\sqrt{\frac{1+2\nu}{1+\nu z}}\Theta\Upsilon^{\frac{1+\nu}{1+\nu z}}. \eeqa
Equating both sides of the inequality Eq. (\ref{r1}), one can find the values of $r_*, r_*'$ that determine
the effective area $\mathcal{S}_*(\Theta,\Upsilon)=\mathcal{S}_*(\mathcal{A})/\Delta^2$ where $v_F/\hat{v}>1$. This leads to the following equations,
\beqa
\label{solrstar}
\zeta_*(\Theta,\Upsilon)=\sqrt{\frac{1+2\nu}{1+\nu z}}\Theta\Upsilon^{\frac{1+\nu}{1+\nu z}}e^{\zeta_*^2(\Theta,\Upsilon)/2},
\eeqa
\beqa
\mathcal{N}_{\rm IKZM}(\Theta,\Upsilon)
=\frac{2\pi}{f}\frac{1+\nu z}{1+2\nu }\Theta^2\Upsilon^{\frac{2\nu}{1+\nu z}}\int_{\{\zeta|v_F>\hat{v}\}}d\zeta \zeta e^{\zeta^2\frac{\nu}{1+2\nu}} ,
\eeqa
which take into account the causality argument and the local dependence of $\hat{\xi}$.
\begin{figure}[t]
\begin{center}
\includegraphics[width=1\linewidth]{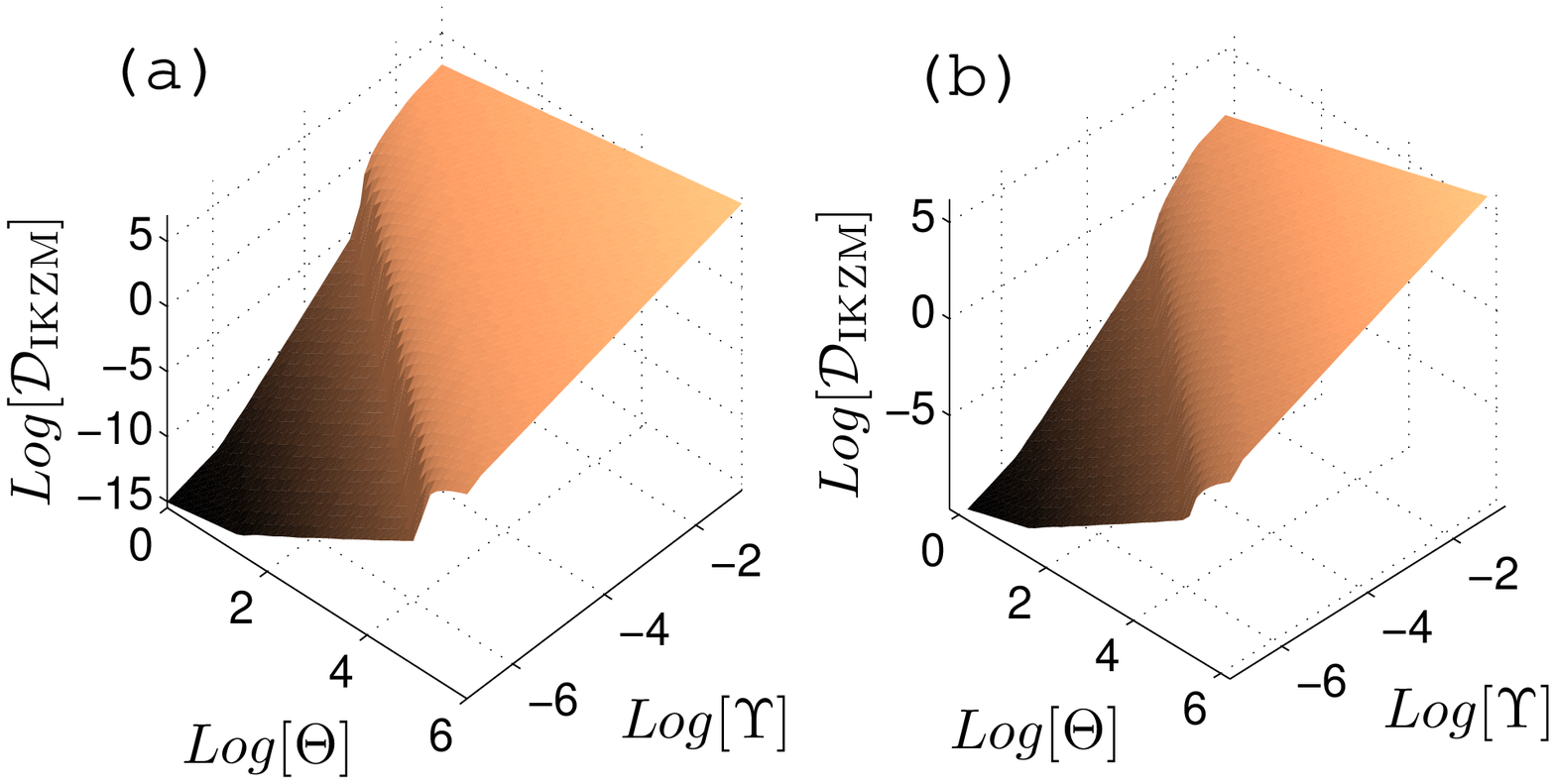}
\end{center}
\caption{\label{figdefecst}
Density of vortices as a function of the dimensionless width of the cloud $\Theta$ and quench rate $\Upsilon$ in an inhomogeneous phase transition
under a linear quench of the temperature a) in a radially symmetric harmonic trap b) in a toroidal trap.
Here $\nu=2/3$, $z=3/2$, but the same qualitative behaviour is observed for $\nu=1/2$, $z=2$.
For small values of ($\Theta$, $\Upsilon$) the density of defects exhibits the IKZM power-law scaling of Eq. (\ref{IKZMscaling}) in a harmonic confinement, and that of Eq. (\ref{IHKZMscaling}) in a toroidal trap. As ($\Theta$, $\Upsilon$) are increased, within the IKZM the power-law dependence breaks down and exhibits a more complicated dependence. For large enough values of $\Theta$, and $\Upsilon$ (such that $\mathcal{A}>\mathcal{A}_c$), nucleation is possible in the whole cloud, the confinement does not play a role any more, an the HKZM scaling in Eq. (\ref{HKZM}) describes the dependence of the density of vortices in both type of traps.}
\end{figure}
Though this expression lacks a simple power-law scaling with the temperature quench and sweeping rate through the transition,
it is still universal,  in the sense that such dependencies are still governed by the critical exponents associated with universality class to which the transition belongs.
Figure \ref{figdefecst}a shows the density of topological defects $\mathcal{D}=\mathcal{N}/\mathcal{S}$ as a function the dimensionless temperature and rate.
As a result of the inhomogeneity the dependence of the density of vortices  on the cooling rate becomes much stronger than in the homogeneous case.
Moreover, the threshold $\mathcal{A}_M$ that arises from the finite-size of the cloud, is responsible for an abrupt jump on density of vortices.
For the sake of illustration in Fig. \ref{figdefecst}, we consider the case in which $n(r_M)/n(0)=0.1$, but the results are qualitatively the same for different values of $r_M$.


In the experiments reported in \cite{BrianAnderson08}, a fitting to the measured temperatures to $(1-{\rm t}/\tau_Q)$ leads to an estimate $\tau_Q\sim 5$ s, while $\tau_0\sim 0.1$ s is given by the scattering time of atoms. Moreover, the de Broglie wavelength is $\xi_0\sim 1.6$ $\mu$m and $\Delta\sim 65$ $\mu$m, so that $\mathcal{A}\sim 1.7$, suggesting a homogeneous scenario in the proximity of the boundary $\mathcal{A}_c$.
Using the explicit form of $\mathcal{A}$ as a guide, we note that an experimental verification of the IKZM, and in particular of the scaling in Eq. (\ref{IKZMscaling}), will be favoured by a slow quench, $\Upsilon=\tau_0/\tau_Q<1$, and a tight trapping potential so as to reduce $\Theta=\Delta/\xi_0$.

We close this section noticing that our discussion is limited to quenches where the correlation length at the freeze out time is much smaller than the trap size. Otherwise, the finite-size scaling modifies the power-law governing the divergence of the correlation length in the neighbourhood of the critical point, as discussed in \cite{FSS}.


\section{Mixing the
mechanisms: Bose-Einstein condensation in a toroidal trap}

The Tucson group also studied the spontaneous nucleation of vortices in a toroidal trap, that we describe in the following.
Such geometry induces a new type of transition  with a mixed homogeneous-inhomogeneous character. To illustrate this, consider a squashed toroidal trap of radius $r_c$, and transverse trapping frequency $\omega_r$.
The corresponding equilibrium density profile of a non-interacting thermal gas has the form $n(\theta, r)=n_0\exp{-(r-r_c)^2/2\Delta^2}$, and as a result the critical temperature becomes
\beqa
T_c(\theta,r)=T_c \exp\Bigg[-\frac{(r-r_c)^2)}{2\Delta^2}\Bigg].
\eeqa
The upshot is that the transition remains homogeneous as a function of $\theta$ but it is governed by causality in the radial direction where it is still inhomogeneous. The relative coordinate $h=r-r_c$ plays the role of $r$ in the preceding section. As in the harmonic trap, a complete suppression of vortex nucleation can arise not only from finite-size effects, but also whenever $v_F<\hat{v}$.
As for the pancake condensate, $\mathcal{A}_c$ governs the transition between the homogeneous and inhomogeneous scenario in the radial direction.
For $\mathcal{A}<\mathcal{A}_c$ the effective size of the cloud is reduced as shown in Fig. \ref{effsize}b, and the IKZM scaling of defects is modified as follows
\beqa
\label{IHKZMscaling}
\mathcal{N}_{\rm IKZM}^{(0)}=\frac{4\pi r_c h_*}{f\hat{\xi}^2}=\frac{4\pi r_c\Delta^2}{f\xi_0^3}
\left(\frac{\tau_02\delta}{\tau T_c(0)}\right)^{\frac{1+3\nu}{1+\nu z}},
\eeqa
under the approximation mentioned above, $h_* =\vert r_*-r_c\vert\ll\Delta$, $\hat{\xi}=\hat{\xi}(r_c)$, which holds for $\mathcal{A}<\zeta(|r_c-r_M|) e^{-\frac{\zeta(|r_c-r_M|)^2}{2}}$.
For the mean-field critical exponents 
$\mathcal{N}_{\rm IKZM}^{(0)}\sim\tau_Q^{-5/4}$,
at variance with the HKZM power law with exponent $1/2$ in Eq. (\ref{HKZM})
 and the inhomogeneous case, see Eq. (\ref{IKZMscaling}) where the exponent is $\frac{2(1+2\nu)}{1+\nu z}=2$.
Similarly, for $\nu=2/3$, $z=3/2$,  the power-law exponent is $3/2$ intermediate between that of HKZM ($2/3$) and IKZM ($7/3$).
The general estimate in the inhomogeneous scenario ($\mathcal{A}>\mathcal{A}_c$) can be obtained by solving numerically  Eq. (\ref{solrstar}) as a function of the dimensionless transverse width of the cloud $\Theta$ and quench rate $\Upsilon$, and noticing that the effective area  is in this case given by
$\mathcal{S}_*(\Theta,\Upsilon)=4\pi \zeta(r_c)[\zeta_*(\Theta,\Upsilon)-\zeta_*'(\Theta,\Upsilon)+\zeta(r_M)]$.

The result is shown in Fig. (\ref{figdefecst})b, which illustrates the change in the scaling from that of Eq. (\ref{HKZM}) (HKZM) to that in Eq. (\ref{IHKZMscaling}) (IKZM) as the quenching time and the inhomogeneity of the cloud are increased. The boundary between the different scalings is now smoother than for the pancake geometry, due to the mixed character of the transition.
We further note that for tight toroidal traps even if vortex nucleation is suppressed, solitons might be formed leading to the spontaneous generation of persistent currents as recently discussed in \cite{Das11}. Nonetheless, for wider traps such solitons become unstable against vortex formation through the snake instability and related mechanisms \cite{Brand}.

\section{Conclusions}
In conclusion, we have shown of the Kibble-Zurek mechanism, extended to describe spatially inhomogeneous systems,
severely modifies with respect to the homogeneous case the scaling of the density of defects as a function of the quenching rate, as illustrated in Table \ref{power_laws_table}.
%
\begin{table}[h]
\begin{center}
\begin{tabular}{c c c c}
\hline
Critical exponents $\setminus$ Trap         &    Homogeneous           &    Harmonic                   &   Toroidal \\
\hline\hline
             &                          &                                &                            \\
Arbitrary    ($\nu$, $z$)
             &  $\frac{2\nu}{1+\nu z}$  &  $\frac{2(1+2\nu)}{1+\nu z} $  &  $\frac{1+3\nu}{1+\nu z}$  \\
             &                          &                                &                            \\
Mean-field theory ($\nu=\frac{1}{2}$, $z=2$)
             &    $\frac{1}{2}$         &      $2$                       &       $\frac{5}{4}$        \\
             &                          &                                &                            \\
Experiments/F model   ($\nu=\frac{2}{3}$, $z=\frac{3}{2}$)
             &     $\frac{2}{3}$        &      $\frac{7}{3}$             &       $\frac{3}{2}$        \\[1ex]
\hline
\end{tabular}
\caption[]{Power laws predicted by the Kibble-Zurek mechanism for the density of vortices $\mathcal{D}$ as a function of the rate $\tau_Q$ of a thermal quench through the critical temperature for Bose-Einstein condensation. The exponent $\alpha$ of the power-law $\mathcal{D}\sim\tau_Q^{-\alpha}$ is shown
for different critical exponents ($\nu$, $z$) and trapping potentials. The scalings in the harmonic toroidal and toroidal trap are restricted to situations where nucleation is limited to a small fraction of the cloud ($r_*,h_*\ll 1$) as discussed in the text.}
\label{power_laws_table}
\end{center}
\end{table}
%
When the presence of a trapping potential leads to a inhomogeneous scenario, a neat power-law scaling governs the nucleation of vortices only when causality limits it to a small fraction of the cloud.
Otherwise, the local dependence of the effective quench rate is to be taken into account, leading to a more complicated behaviour, different from a power-law.
We have further introduced a simple parameter $\mathcal{A}$ which allows one to estimate which is the relevant scenario for the nucleation of defects in a trapped cloud. Indeed, the standard Kibble-Zurek scaling is recovered for weak trapping potentials and fast quenches.
We close noticing that while the homogeneous Kibble-Zurek mechanism has been studied in a wide variety of experiments \cite{kzmexp},
the mechanism for the nucleation of topological defects in inhomogeneous systems lacks to date experimental evidence.
Hence, we hope that measurements of the number of spontaneously generated vortices
as a function of the cooling rate of an atomic cloud undergoing Bose-Einstein condensation, might soon change this state of affairs.

{\it Acknowledgements.}
This work was motivated by a talk by and discussions with Matt Davies at the BEC2009.
AdC acknowledges Marvin D. Girardeau and Brian P. Anderson  for discussions and hospitality in the College of Optical Sciences at
the University of Arizona. Further comments by W. H. Zurek, J. Dziarmaga, R. Rivers and E. Vicari are gratefully acknowledged, as well as support from EPSRC project
(EP/E058256), an Alexander-von-Humboldt Professorship and the EU STREP Projects HIP and PICC.

\end{document}